# A Direct Real-Time Observation of Anion Intercalation in Graphite Process and Its Fully Reversibility by SAXS/WAXS Techniques

*Giorgia Greco,\* Giuseppe Antonio Elia,\* Daniel Hermida-Merino, Robert Hahn, and Simone Raoux*

**The process of anion intercalation in graphite and its reversibility plays a crucial role in the next generation energy-storage devices. Herein the reaction mechanism of the aluminum graphite dual ion cell by operando X-ray scattering from small angles to wide angles is investigated. The staging behavior of the graphite intercalation compound (GIC) formation, its phase transitions, and its reversible process are observed for the first time by directly measuring the repeated intercalation distance, along with the microporosity of the cathode graphite. The investigation demonstrates complete reversibility of the electrochemical intercalation process, alongside nano- and micro-structural reorganization of natural graphite induced by intercalation. This work represents a new insight into thermodynamic aspects taking place during intermediate phase transitions in the GIC formation.**

## 1. Introduction

Nowadays, storing and converting "clean" energy is mandatory due to the increasing global demand for renewable energy, thus helping to mitigate climate change caused by burning fossil fuels.[1] For this reason, the mechanism of intercalation, more in general, the property of some solids to host atoms or molecules, became of great interest not only because our knowledge of the mechanism is still incomplete,[2] but also because of practical application in high-energy-density batteries and hydrogen-storage systems. Since the 1940s,[3,4] already in the initial studies on intercalation phenomena in graphite, X-ray diffraction has been the most used technique to indirectly observe the structural modifications of graphite induced by the intercalating species which follows a staging mechanism.[5–10] An indirect measure needs the use of models and theoretical calculations to reconstruct the current intercalation stage of the graphite intercalation compound (GIC).[8,10] A direct observation instead allows the estimation of the CIG formation process without the need for simulations or theoretical calculations. In particular, small angles (SAXS) can be a very suitable methodology for evaluating the GIC formation as it can directly detect the periodic repeat distance $I_c$, allowing the direct evaluation of staging and its intermediates phase transitions with high accuracy.[11,12]

G. Greco[+], S. Raoux
Helmholtz–Zentrum Berlin für Materialien und Energie GmbH
Hahn–Meitner–Platz 1, D–14109 Berlin, Germany
E-mail: giorgia.greco@uniroma1.it

G. Greco[+]
Chemistry Department
Sapienza University of Rome
P.le Aldo Moro 5, Roma 00185, Italy

G. A. Elia[++], R. Hahn
Technical University of Berlin
Research Center of Microperipheric Technologies
Gustav–Meyer–Allee 25, D–13355 Berlin, Germany
E-mail: giuseppe.elia@polito.it

D. Hermida-Merino[+++]
DUBBLE-Dutch Belgian Beamline (BM26)
ESRF, 6 Rue Jules Horowitz
BP 220, 38043, Grenoble CEDEX 9, France

R. Hahn
Fraunhofer IZM
Institut für Zuverlässigkeit und Mikrointegration
Gustav–Meyer–Allee 25, D–13355 Berlin, Germany

[+]Present address: Department of Fusion and Technology for Nuclear Safety and Security, ENEA Centro Ricerche Casaccia, Via Anguillarese 301, Rome 00123, Italy

[++]Present address: Department of Applied Science and Technology (DISAT), Politecnico di Torino, Corso Duca degli Abruzzi 24, Torino 10129, Italy

[+++]Present address: Departamento de Física Aplicada, CINBIO, Universidade de Vigo, Campus Lago-as-Marcosende, Vigo, Galicia 36310, Spain

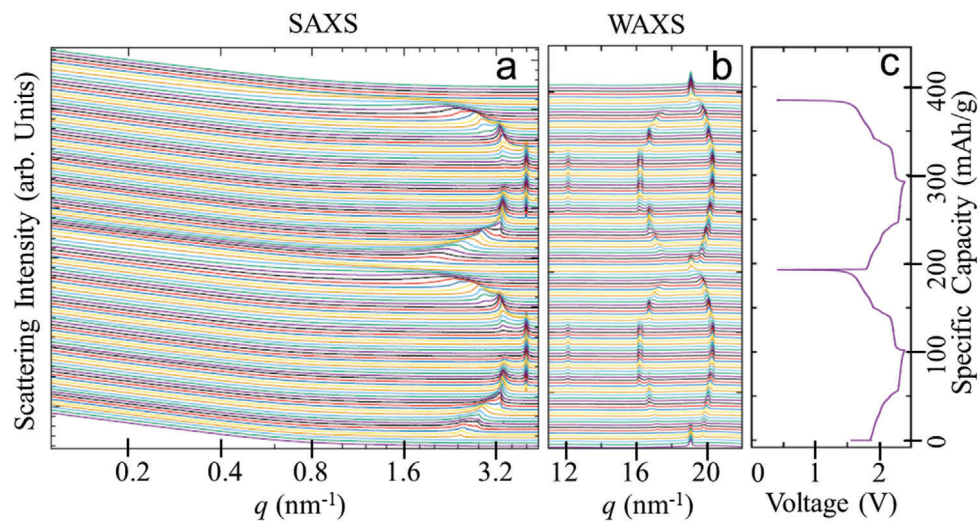

**Figure 1.** Operando study of aluminum battery in the first two cycles. On the left side (a), the small-angle range is plotted in logarithmic scale, the central panel (b) shows the wide-angle range, and in the right panel (c), the actual potential curve acquired during the experiment. To get a better and complete overview of the process, the SAXS and WAXS patterns were shifted throw the y axis.

More recently, the direct observation of GIC formation was observed by a combination of RAMAN and optical microscopy.[2,13] However, those techniques are surface sensitive and could only give partial information about the bulk transition. In 2018, Pan et al.[14] showed the direct measurements of the intercalated repeated distance in the first charge of aluminum graphite dual ion cells (AGDICs). The AGDICs system is based on the anion ($AlCl_4^-$) intercalation between graphite layers[7,12–15] and have attracted significant interest as a potential electrochemical storage system characterized by high power and elevated cycle life.[16–24] The further understanding of the reaction mechanism can provide not only development on AGDICs but also for application in Li, Na, K, Mg, and Ca batteries, and hybrid systems.[7,18]

Anion intercalation phenomenon into graphite at the positive electrode is a very versatile electrochemical storage mechanism that can be coupled with many negative electrode systems, from the more popular lithium[25,26] to more recent alternatives such as Na,[27] K,[28] Mg,[29] Ca,[30] and Al[31] Moreover, GICs formation mechanism is extremely relevant for cation intercalation, using graphite as the negative electrode, for alkali metals such as Li[32] and K[33] as well as for Na intercalation mediated through solvent cointercalation.[34]

The characteristics of the AGDICs make them particularly suitable for applications in stationary storage systems, where long cycle life and low cost are highly desirable.[19,22,24,35] Among the diverse types of natural and synthetic graphite that have been investigated for application in AGDICs, natural graphite (NG) has shown the most attractive characteristics in terms of the delivered capacity, cycle life, and power density.[17,18,36,37] Due to its good performance and long-term stability, this system represents an interesting and suitable model to study the intercalation mechanism in NG.

Herein we carried out a detailed structural characterization of the Al/1-ethyl-3-methylimidazolium chloride (EMIMCl):aluminum trichloride ($AlCl_3$)/NG system, evaluating its reversibility by operando X-ray scattering from SAXS to wide angles (WAXS). The characterization allows for determining the relationship between the structural modifications of the NG cathode and the electrochemical behavior, giving a more comprehensive understanding of the intermediated phase transitions during GIC formation, an aspect that is not still clear.[2,10,38]

## 2. Results and Discussion

**Figure 1** shows the SAXS and WAXS spectral evolution during galvanostatic cycling of the AGDIC. The voltage signature of the galvanostatic cycling performed during spectral acquisition is reported along with the spectra that were collected simultaneously during cycling. Figure S1, Supporting Information, displays a picture of the set-up and the operando cell.

At a glance, Figure 1 gives a complete overview of the graphite electrode's structural changes upon cycling, indicating a structural evolution that goes through various GIC intermediates.[8,10,39] The voltage profile reported in Figure 1c shows an initial sloping voltage profile between 0 and about 50 mAh g$^{-1}$, where a shift of the peak to higher $q$ values is observed in the SAXS spectra (Figure 1a). The second part of the voltage profile, from about 50 mAh g$^{-1}$ until the end of the charging process, has a flat voltage profile (Figure 1c), corresponding to the gradual disappearance of the peak located at $q \approx 3.2$ nm$^{-1}$, and the appearance of a new peak at $q = 4$ nm$^{-1}$. Similar behavior is also present in the WAXS spectra reported in Figure 1b, where the (002) reflection of graphite disappears alongside the formation of two new reflections associated with the formation of the GIC. These reflections gradually shift during the initial phase of the charging process until a capacity of about 50 mAh g$^{-1}$ is reached. The peak at $q = 16.5$ nm$^{-1}$ can be observed to disappear progressively alongside the formation of two new peaks at $q = 16$ and 12 nm$^{-1}$.

**Figure 2** reports selected spectra obtained during the operando experiment highlighting phase transitions between stages. At open circuit voltage (OCV), the pattern reveals the typical main peak (002) at $q \approx 18.7$ nm$^{-1}$ of the NG.[11,22] As soon as the current

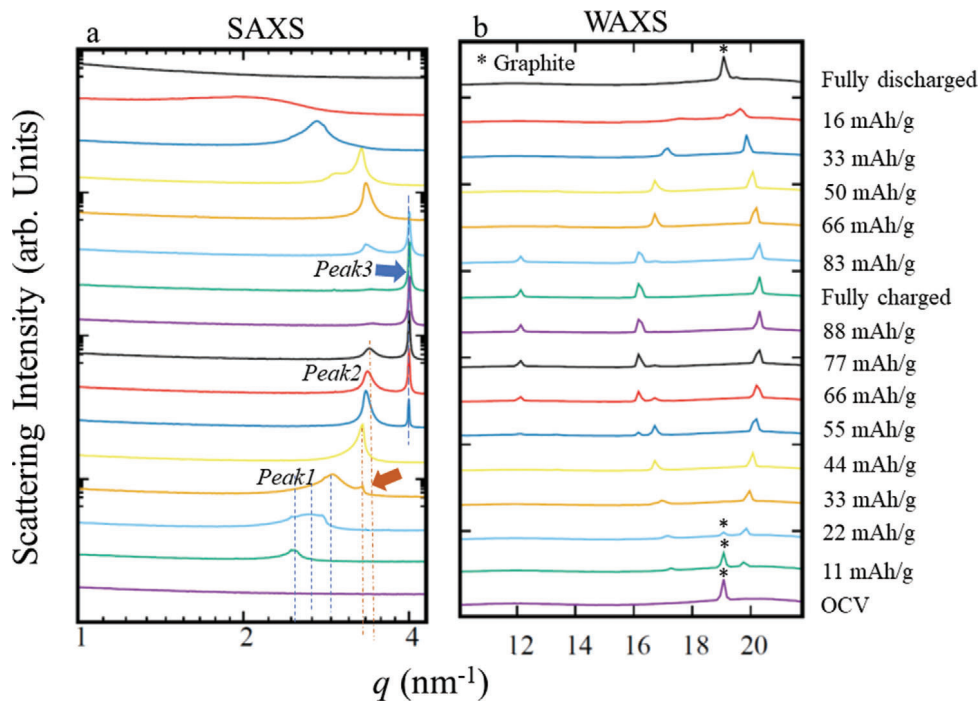

**Figure 2.** Selected a) SAXS and b) WAXS profiles of the first cycle along with the relative capacity. The arrows in (a) indicate the development of a new peak corresponding to a phase transition. The star in the WAXS profiles, shown in (b), indicate the (002) peak of graphite.

(25 mA g$^{-1}$) starts to flow, a smooth peak appears in the SAXS pattern at $q = 2.6$ nm$^{-1}$, which we refer to as Peak1. In the WAXS region, two less intense peaks appear, and the main peak of the graphite can be observed to decrease (see Figures 1 and 2). It has already been demonstrated[11,12,14] that the peak in the SAXS region is related to the intercalated AlCl$_4^-$ species, and in particular is the intercalated repeated distance, which is defined as:

$$I_c = \Delta d + n \cdot 0.335 \text{ nm} \quad (1)$$

where $\Delta d$ is the intercalant gallery height, $n$ is the number of free graphene layers, and 0.335 nm is the distance between them.[11,22] The peaks that arise in the WAXS range come from Bragg reflections.[40] In the $q$ range, we have:

$$I_c \approx 2\pi/q_c \quad (2)$$

and the additional reflections:

$$q_c \approx m \cdot q \quad (3)$$

where $q_c$ is the peak position in the SAXS region, $m$ is an integer equal to 1, 2, 3..., and $q$ is the momentum transfer or scattering vector.[14,41] The peaks appear for $m = 5$ and 6 in the WAXS range in correspondence with the first voltage plateau at 50 mAh g$^{-1}$ in Figure 2. The other integer reflections are not evident (see Figures 1 and 2). This is due to the weak intensity of the peaks and the disordered arrangement of the intercalant species within the host structure, as has also been observed in the work of Pan et al.[42] As charging is continued, Peak1 in the SAXS region shifts continuously toward higher $q$ values, and the peak becomes sharper. At the same time, the signal at $q \approx 2.4$ nm$^{-1}$ starts to disappear alongside the appearance of a sharper peak at $q \approx 2.6$ nm$^{-1}$, which we refer to as Peak2, which is highlighted in Figure 2a with the orange arrow. In the corresponding WAXS patterns, this first phase transition is less evident, but the graphite peak at $q \approx 18.7$ nm$^{-1}$ begins to disappear while the two additional peaks increase. This behavior indicates an ordering of the intercalant species arrangement in the host structure; see Figures 1b and 2b. As can be seen in Figure 1c, when the capacity reaches ≈30 mAh g$^{-1}$, a small bump in the voltage curve is present, which coincides with the appearance of Peak2.

Between ≈30 and 50 mAh g$^{-1}$, Peak2 shifts slightly (see Figure 2a and at ≈50 mAh g$^{-1}$, a well-defined peak at ≈4 nm$^{-1}$ appears, which we refer to as Peak3. This peak is highlighted by the blue arrow in Figure 2a. During the second voltage plateau between 50 and 100 mAh g$^{-1}$ (Fully charged in Figure 2), two well-defined GIC phases coexist. In the WAXS range (Figures 1b and 2b), four peaks are present, which correspond to $m = 3, 4, 5$, and 6 Bragg reflections. Lower integer reflections become evident due to the higher ordering degree of the intercalant anions. Peak3 remains at a position of ≈4.0 nm$^{-1}$ and increases in intensity until Peak2 almost disappears (see Figure 2), but it is nevertheless still present with low intensity in the fully charged state.

The battery is fully charged at 100 mAh g$^{-1}$, and a single, sharp, and intense peak is evident at $q = 4$ nm$^{-1}$, corresponding to a stage of $n = 3$ from theoretical calculations,[42] see Figures 1 and 2.

Figures 1 and 2 show that the discharge behavior of the battery is perfectly symmetric to the charging behavior in both the SAXS and WAXS ranges. When the battery is fully discharged, the peak of the intercalant gallery distance $I_c$ disappears in the SAXS range along with the appearance of the (002) peak of graphite in the WAXS range, thereby revealing the full reversibility of anion release (Figures 1 and 2).

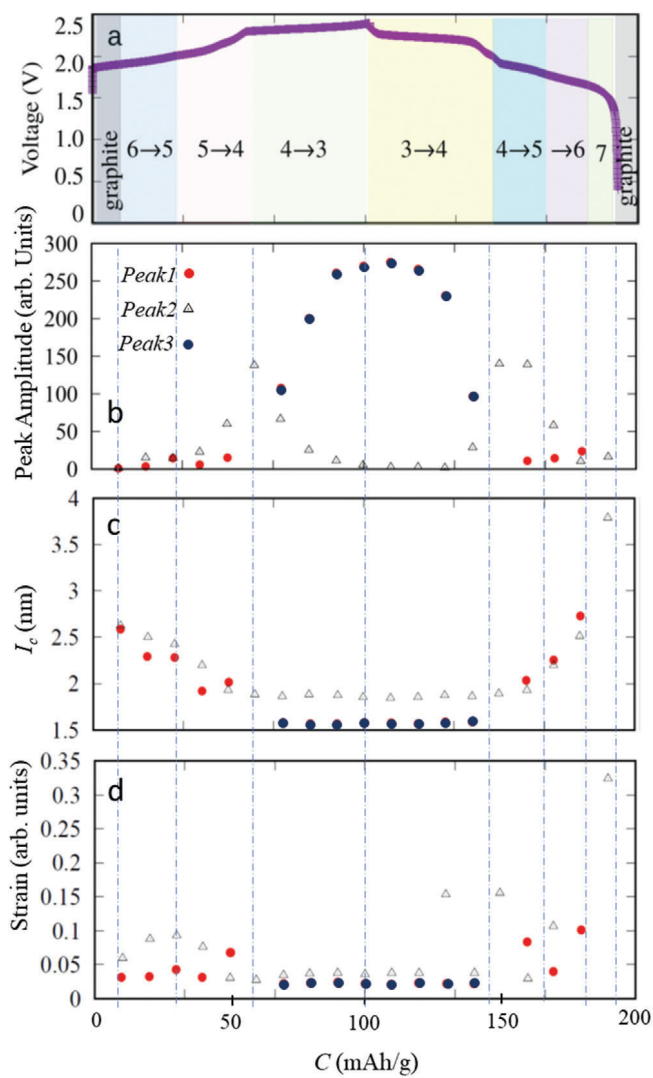

**Figure 3.** SAXS fitting parameter results along with the electrochemical performance of the operando NG/Al cell. The voltage profile acquired during the experiment (a), the intercalant gallery distance $I_c$ (Equation (1)) obtained from the SAXS peaks positions (Peak1, Peak2, and Peak3) (b), the SAXS peak amplitudes (c), and peak strain behavior as a function of time (d). In the inset of (a), the numbers indicate the stages and the staging transitions upon cycling.

Figure S2, Supporting Information, reports patterns collected on the bare electrolyte during the first cycle, evidencing no observable changes, suggesting the negligible contributions of the electrolyte over the full SAXS range.

**Figures 3** and **4** show the results obtained by fitting the SAXS curves with a model represented by the sum of four contributions: background, microporosity (pore size < 1 µm diameter), and two Gaussian peaks,[41] the details of the model are described below. All SAXS profiles have been fitted with the same model. Figure S3, Supporting Information, reports an example of the fit for the 30 mAh g$^{-1}$ sample. Figure 3 summarizes the fit parameters obtained as a function of capacity and the voltage curve of the cell (Figure 3a). Figure 3b shows the amplitude of peaks related to $I_c$ shown in Figure 3c. Figure 3d shows the strain calculated by the formula $\Delta q/q$.[41]

Additionally, the WAXS profiles were fitted as described below. The fit results are shown in the Supporting Information. The strain behavior and peak position obtained agree with the SAXS results and the Bragg reflections.

The peak evolution reported in Figure 3 shows that two broad and low-intensity peaks overlap in Peak1 and appear just after the current begins to flow in the cell. The peak positions and broadening suggest the occurrence of mixed staging[43] regions around stage 6. With an increasing state of charge, $I_c$ gradually decreases, and the strain starts increasing up to ≈30 mAh g$^{-1}$ in correspondence with a first phase transition. In this region, we can assume that upon anion intercalation, the staging continuously shifts from 6 to 5.

In the second half of the voltage plateau, the intensity of Peak2 increases until ≈50 mAh g$^{-1}$, where it is possible to distinguish a single sharp peak. At this point, the strain associated with the most intense peak decreases, and the staging evolves from 5 to 4. The complete transition from stage 5 to 4 corresponds with the inflexion of the voltage and the start of the flat voltage plateau. During this charging stage, only Peak2 is present in the SAXS region.

The second voltage plateau is characterized as a two-stage region,[43] where two well-defined peaks (Peak2 and Peak3) are evident in the SAXS region, indicating a transition between stages 4 and 3, see Figures 2 and 3b. In this second phase of the charging process (≥50 mAh g$^{-1}$), the position and strain associated with the peaks remain approximately constant, but there is a rapid increase in the amplitude of Peak3, which is associated with stage 3. This persists until the battery is fully charged (100 mAh g$^{-1}$). As already described above, the discharge behavior is highly symmetric with the charging process. One slight deviation from symmetry is evident at the end of the discharge process, where the strain increases and $I_c$ shows a broad peak proper of stage 7, which has not previously been observed (see Figures 2 and 3c,d). This phenomenon can be associated with nano- and micro-structural reorganization of NG induced by the intercalation during the first cycle.

The volume size distribution of the micropores of the NG for different states of charge was also evaluated to investigate changes in porosity, as reported in Figure 4. It can be seen that there is a certain degree of microporosity in NG at OCV, which decreases upon charging. The $AlCl_4^-$ ions intercalate through the NG lattice, generating a volumetric expansion of the cathode and thereby decreasing the porosity,[11,39] which almost disappears at the fully charged state. The release of $AlCl_4^-$ induces cathode microporosity, which grows during discharge (Figure 4). After full discharge, the microporosity increases by three times relative to the initial microporosity. Figure S4, Supporting Information, compares the pore size distribution of the electrode at the beginning and after the second cycle.

**Figure 5** compares the SAXS and WAXS profiles of the electrode before and after two cycles. The slope of the SAXS curve at OCV for lower $q$ values follows the expected trend of $q^{-4}$,[11,22] indicating a smooth surface of the NG crystallites.[11,22] A com-parison of the SAXS slopes reported in Figure 5a indicates an increase in the overall scattering intensity, which is most likely associated with a difference in contrast that can be attributed to a homogeneously distributed increase in porosity, as is also implicated in Figure 4. In the SAXS region, the increase in intensity at low $q$ values can be attributed to crystallite form factor orientation changes, as confirmed by the changes in WAXS peak intensities before and after two cycles. Figure 5b demonstrates a total recovery of the graphite peaks but a change in the relative intensities of the peaks (100) and (101). This behavior indicates a preferential orientation of the crystal lattice during the insertion of the intercalant and an associated structural reorganization. The 2D SAXS images reported in Figure S5, Supporting Information, demonstrate amplitude intensity modulations, which are characteristic of the partial orientation of the sample.[44]

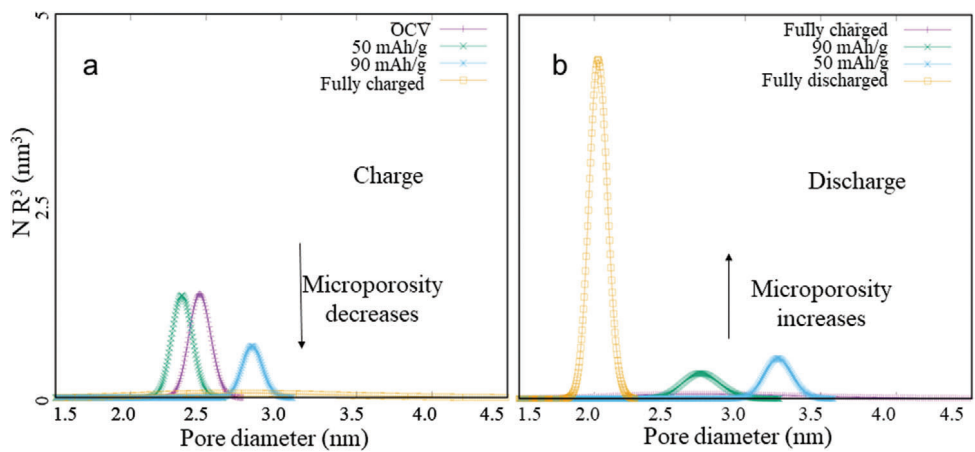

**Figure 4.** Volume weighted size distribution of micropores in the electrodes at different states of charge. (a) reports the microporosity during charge and (b) reports the mesoporosity during discharge, with the arrows indicating the trends in changing microporosity.

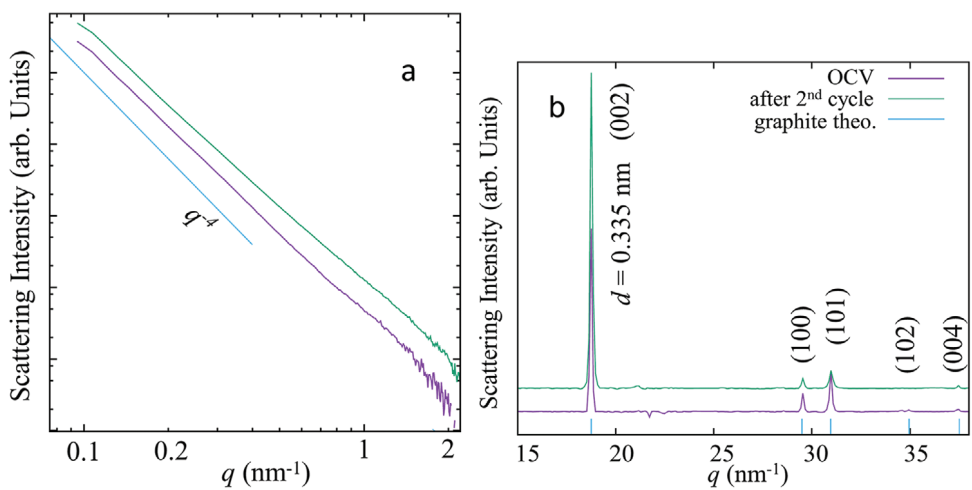

**Figure 5.** a) SAXS and b) WAXS profiles after background subtraction of the electrode at OCV and in the fully discharged state after the second cycle.

**Figure 6** shows an overall schematic of the $AlCl_4^-$ intercalation process based on the model proposed by L. B. Ebert.[45] In this work, we focus on the meso- and micro-structure of graphite (range below µm), so we are not sensitive to the blister formation characterized and modeled by Hathcock and Murray in 1995.[9] As the current begins to flow through the battery, the $AlCl_4^-$ anions randomly intercalate through preferential graphite planes, leading to a combination of stages, including pure graphite, as depicted in Figure 6b. The intermediate and disordered phases disappear when the battery is half-charged, and a more well-defined stage emerges, as shown in Figure 6c. When the battery is fully charged, the second phase transition from stage 4 to stage 3 ends; at this stage, the degree of order is very high, and it is likely that the intercalated species further arranged themselves, as depicted in Figure 6d.

## 3. Conclusions

In this study, operando SAXS/WAXS characterization during electrochemical intercalation of $AlCl_4^-$ into NG was used to provide new insight into the anion intercalation process in graphite electrodes. The study demonstrated the complete electrochemical and structural reversibility of the anion intercalation process, following the GIC staging process in detail.

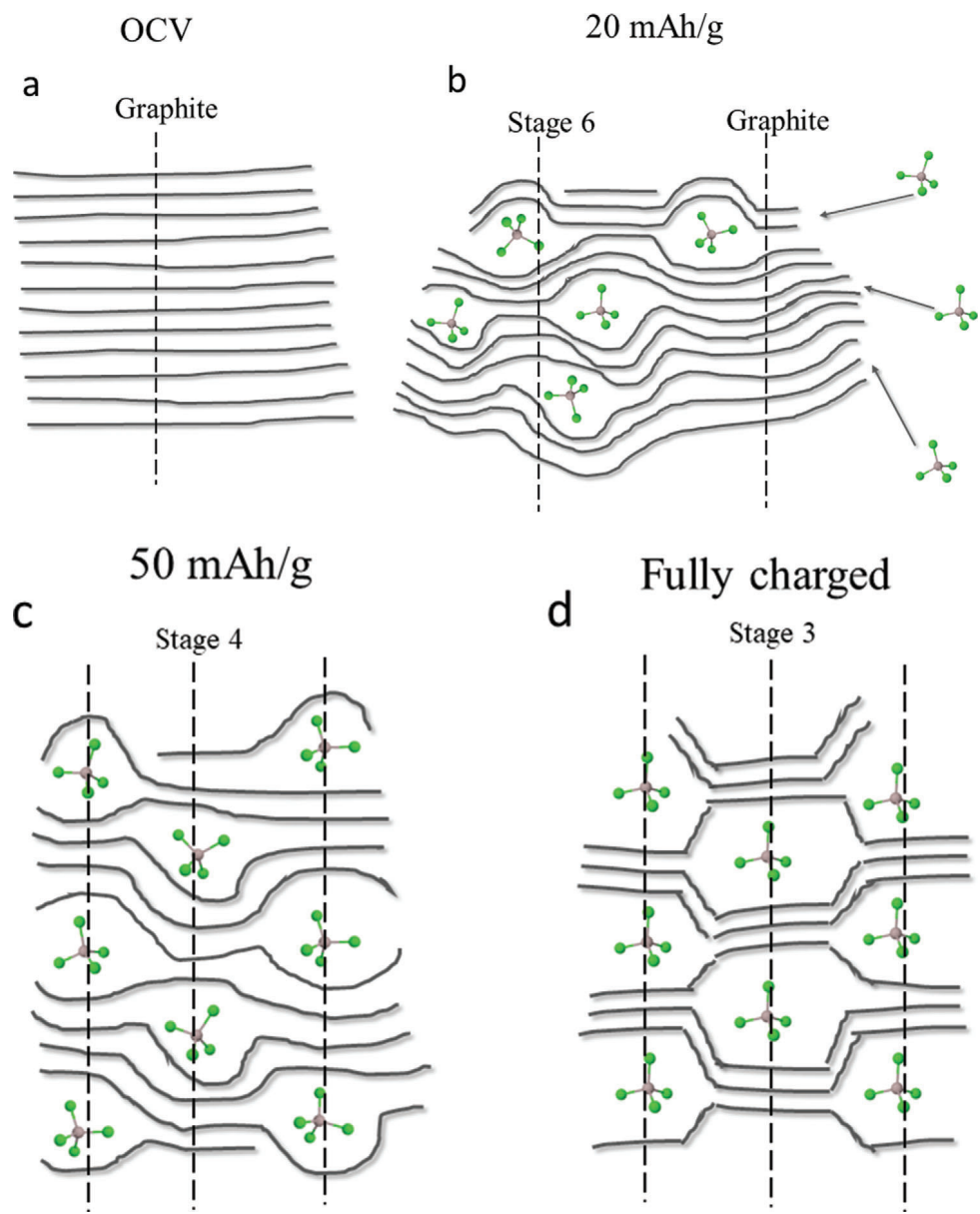

**Figure 6.** Schematic view of the intercalation process at selected states of charge: at OCV (a), at the beginning of the intercalation process and corresponding to the beginning of the first voltage plateau (b), when the first phase transition to stage 4 is complete (c), and when the battery is fully charged and a well ordered stage 3 is present (d).

The operando investigation allowed a detailed mapping of the GIC staging behavior during the electrochemical process, which demonstrated a disordered intercalation characterized by mixed staging in the initial phase of the process, followed by a more ordered intercalation process between stage 4 and stage 3 GICs. Additionally, the characterization revealed a nano- and microstructural reorganization of NG induced by the intercalation in the first cycle, with an associated increase in the electrode microporosity. A more in-depth comprehension of the anion intercalation process represents a step forward in the development of this class of electrochemical energy storage systems. In addition, this system also has a broader interest in basic research of ion intercalation phenomena and GIC formation.

## 4. Experimental Section

*Sample Preparation and Operando Cell Assembly*: The electrolyte EMIMCl:AlCl$_3$ in a 1:1.5 molar ratio was provided by IOLITEC, with a water content of the electrolyte lower than 100 ppm. The NG powder used as the cathode material was provided by PLANO GmbH.[17] The NG electrode was produced by spray deposition on a Whatman GF/A glass fiber separator.[17,46] The NG electrodes employed in these tests had an active material loading of 5 mg cm$^{-2}$. The electrochemical measurements were performed using PEEK self-made cells. Figure S1b, Supporting Information, shows a picture of the cell used for the test. The cycling tests of the Al/EMIMCl:AlCl$_3$/NG cell was carried out by applying a specific current of 25 mA g$^{-1}$ in the voltage range 0.4–2.4 V. This cell configuration was selected for its stability against the highly corrosive EMIMCl:AlCl$_3$ electrolyte.[12,17,47,48]

*Operando Set-Up*: SAXS and WAXS measurements were performed at the European Synchrotron Radiation Facility at the beamline bm26 B station, DUBBLE SAXS/WAXS beamline,[49] Grenoble, France. A picture of the set-up is shown in Figure S1, Supporting Information. The beamline could generate a photon flux of $\approx 10^{11}$ photons s$^{-1}$ in a 300 × 300 µm focused beam using sagittal and bent mirror focusing optics. The photon energy was used to collect simultaneously SAXS and WAXS measurements with an energy of 12 keV. In order to have the same sensitivity for WAXS and SAXS, the same detector technology was selected. The SAXS pattern was collected using a Pilatus 1M detector with a sensitive area of 168.7 × 179.4 mm$^2$. The WAXS pattern was collected with a linear Pilatus 300 kw detector with a sensitive area of 253.7 × 33.5 mm$^2$.[50] For both detectors the pixel size was 172 × 172 µm. The absence of readout noise and dark current, a sharp point spread function, and a high dynamic range (approximately one million counts per pixel) allowed the acquisition of high-quality data even at very short exposure times. The measurements were acquired for 5 s every 10 min. The SAXS $q$ range went from 0.1 to 4.3 nm$^{-1}$, between 62.8 and 1.4 nm.

*SAXS/WAXS Data Reduction*: The 2D SAXS and WAXS images were integrated in the DUBBLE[51] program and 1D profiles were obtained.

*Background Subtraction and SAXS Fit*: The electrolyte contribution, shown in Figure S2, Supporting Information, was first subtracted from the total scattering curve of all the curves and in particular for Figure 3. The fluorescence scattering or background was also subtracted by the Porod plot,[44] which consisted of plotting the SAXS curve as $y = q^4 \Delta I(q)$ versus $x = q^4$. It was then possible to obtain a straight line defined as $q^4 \Delta I(q) = K_p + q^4$, where $K_p$ represented the intercept of the straight line and the so-called fluorescence background. To remove the fluorescence background, it was sufficient to remove the obtained $K_p$ for all scattering curves[44] after the background subtraction, each peak, labeled as Peak1, Peak2, and Peak3, were fitted with a Gaussian function with the SASfit[52] program and the micropores were considered spherical as first approximation. The Gaussian function was chosen in order to direct calculate the strain by the formula $\frac{\Delta q}{q}$.[41] The results of which are shown in Figure 3.

*WAXS Data Treatment and Fit*: The WAXS patterns were fitted by Fityk[53] program. After the background subtraction, the peaks were fitted by a Gaussian function in order to compare SAXS and WAXS results. The obtained WAXS results are in the Supporting Information.

## Supporting Information

Supporting Information is available from the Wiley Online Library or from the author.

## Acknowledgements

The authors would like to thank Klaus Effland for his contribution in the operando cell design and construction. The authors would also like to thank Marko Perestjuk and Josè Fernando Queiruga Rey for their help during the beamtime. The European Commission funded this research within the H2020 ALION project under contract 646286 and the German Federal Ministry of Education and Research in the AlSiBat project under contract 03SF0486 and the project ALIBATT under contract 03XP0128E. D.H.-M acknowledges the "Maria Zambrano" contract for the University of Vigo, financed by the Spanish Ministerio de Universidades/33.50.460A.752 and by "European Union NextGenerationEU/PRTR". This project has received funding from the European Union's Horizon 2020 research and innovation programme under the Marie Sklodowska-Curie grant agreement No 101029608 and "Progetti Ateneo" project no. AR2221815D2AD501.

Open Access Funding provided by Universita degli Studi di Roma La Sapienza within the CRUI-CARE Agreement.

## Conflict of Interest

The authors declare no conflict of interest.

## Data Availability Statement

The data that support the findings of this study are openly available in SAXS/WAXS operando on Al dual battery at ESRF at https://doi.org/10.17632/sp98v4nvh7.1, reference number 17632.

# CORRECTION

# A Direct Real-Time Observation of Anion Intercalation in Graphite Process and Its Fully Reversibility by Saxs/Waxs Techniques

*Giorgia Greco, Giuseppe Antonio Elia, Daniel Hermida-Merino, Robert Hahn, and Simone Raoux*

*Small Methods* **2023**, *7*, 2201633

DOI: 10.1002/smtd.202201633

The funding information is being corrected from

"The authors would like to thank Klaus Effland for his contribution in the operando cell design and construction". The authors would also like to thank Marko Perestjuk and Josè Fernando Queiruga Rey for their help during the beamtime. The European Commission funded this research within the H2020 ALION project under contract 646286 and the German Federal Ministry of Education and Research in the AlSiBat project under contract 03SF0486 and the project ALIBATT under contract 03XP0128E. D.H.-M acknowledges the "Maria Zambrano" contract for the University of Vigo, financed by the Spanish Ministerio de Universidades/33.50.460A.752 and by "European Union NextGenerationEU/PRTR".

Open Access Funding provided by Universita degli Studi di Roma La Sapienza within the CRUI-CARE Agreement."

to

"The authors would like to thank Klaus Effland for his contribution in the operando cell design and construction". The authors would also like to thank Marko Perestjuk and Josè Fernando Queiruga Rey for their help during the beamtime. The European Commission funded this research within the H2020 ALION project under contract 646286 and the German Federal Ministry of Education and Research in the AlSiBat project under contract 03SF0486 and the project ALIBATT under contract 03XP0128E. D.H.-M acknowledges the "Maria Zambrano" contract for the University of Vigo, financed by the Spanish Ministerio de Universidades/33.50.460A.752 and by "European Union NextGenerationEU/PRTR". **This project has received funding from the European Unions Horizon2020 research and innovation programme under the Marie SklodowskaCurie grant agreement no. 101029608**

Open Access Funding provided by Universita degli Studi di Roma La Sapienza within the CRUI-CARE Agreement."

The authors and the editorial office sincerely apologize for any inconvenience caused by this correction.

DOI: 10.1002/smtd.202300769

# small methods

## Supporting Information

A Direct Real-Time Observation of Anion Intercalation in Graphite Process and Its Fully Reversibility by SAXS/WAXS Techniques

*Giorgia Greco\*, Giuseppe Antonio Elia\*, Daniel Hermida-Merino, Robert Hahn and Simone Raoux*

# Supporting Information

**An operando study on the full reversibility of aluminium graphite dual ion cells**

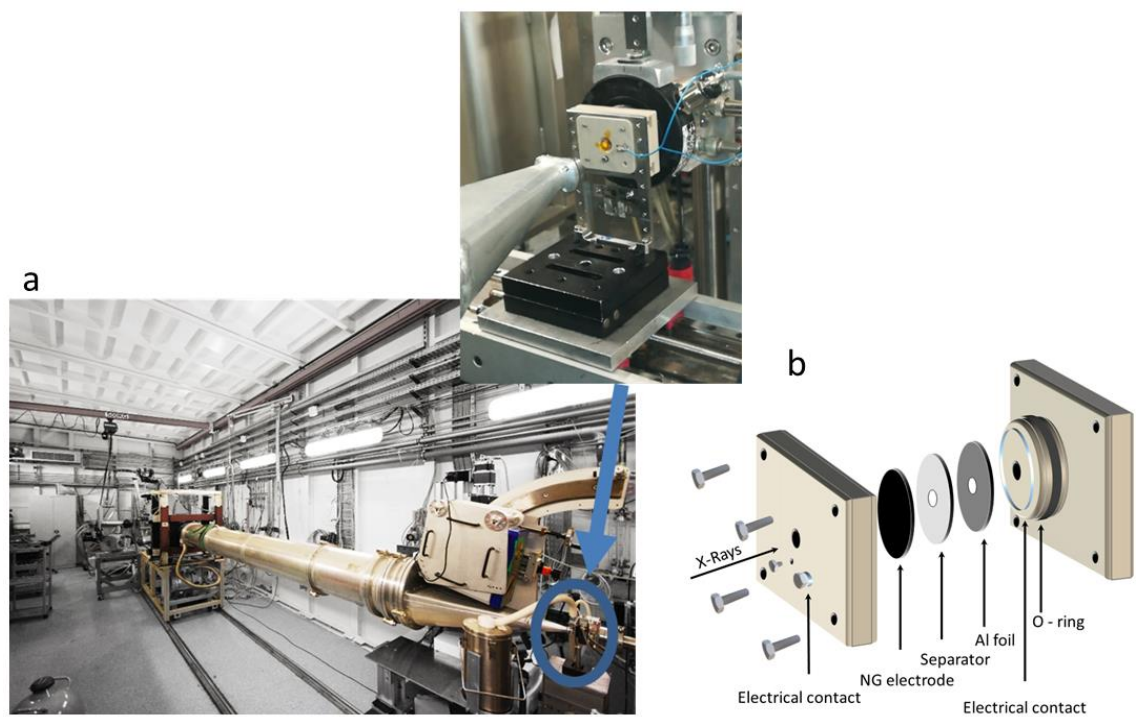

**Figure S1** A picture of the set-up and an enlargement of the operando cell used at the DUBBLE beamline (a). The schematic view of the operando cell used (b).

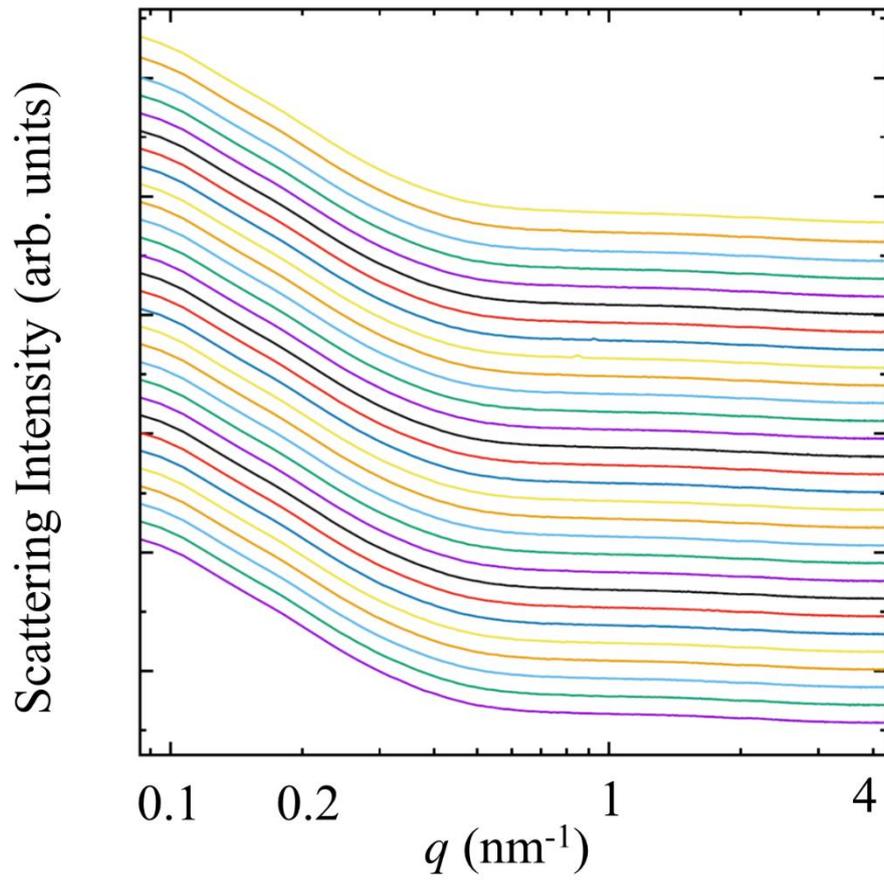

**Figure S2** SAXS profile of the electrolyte at the first cycle, measurements were collected every 10 min.

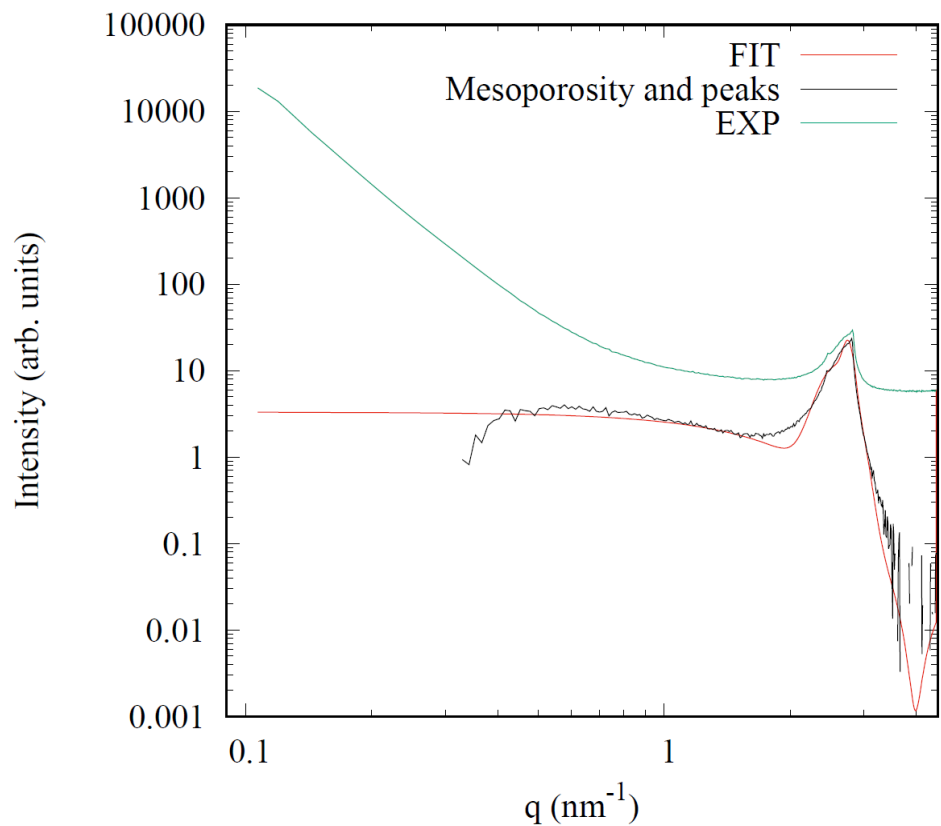

**Figure S3** Best fit (red line) of the SAXS profile of the natural graphite electrode after 30 minutes, the lower curves represents the contribution of microporosity (represented by spheres in first approximation) and two peaks (see also **Figure 3**) after the background subtraction (black line). The upper curve represents the total scattering curve (greenline).

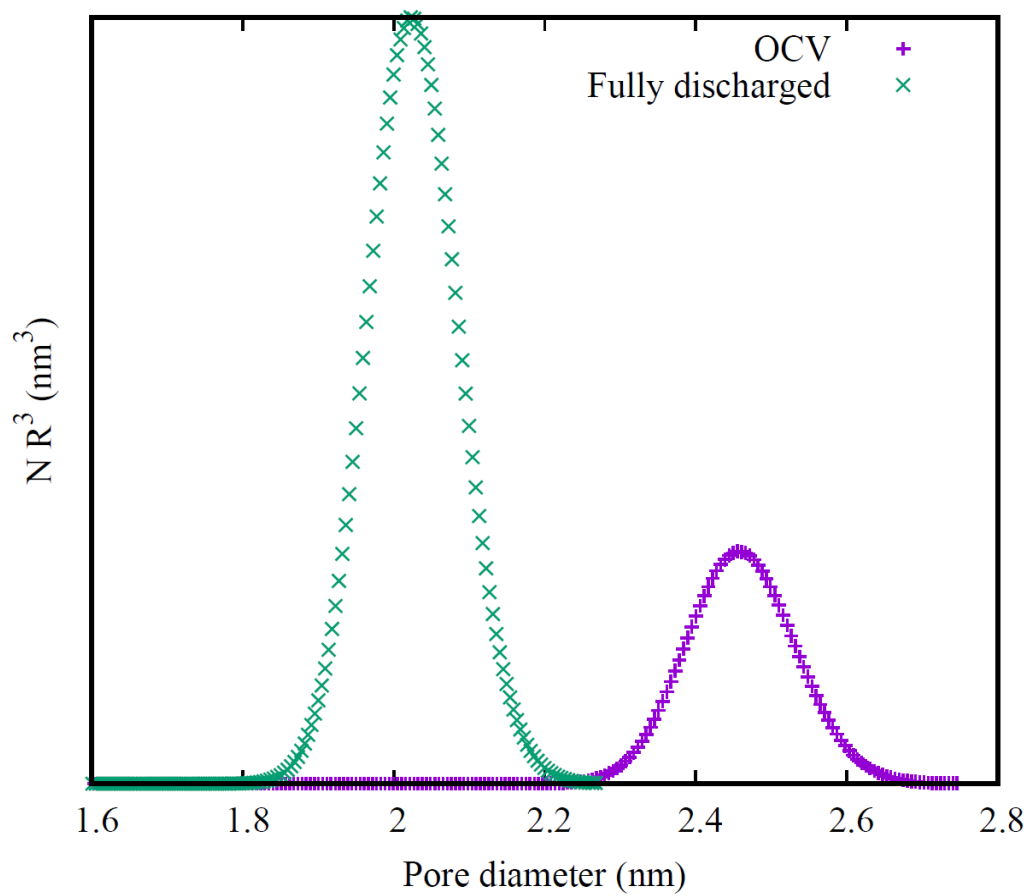

**Figure S4** Comparison of pore size distribution of the same electrode before and after 2 cycles.

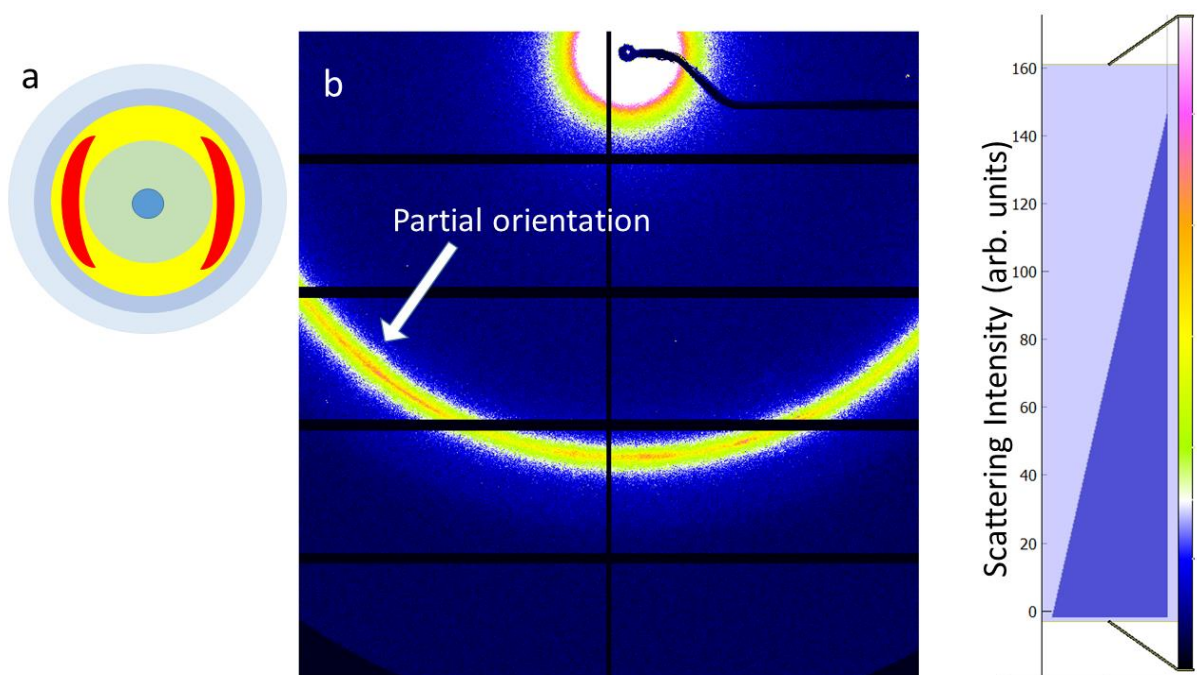

**Figure S5** Schematic representation of the 2D SAXS image of a partially oriented sample (a). A SAXS 2D image collected upon cycling ant its scattering intensity panel, with an indication of partial orentation highlighted by the arrow (b).

4